\documentclass[sigconf]{acmart}
\AtBeginDocument{%
  }

\copyrightyear{2026}
\acmYear{2026}
\setcopyright{cc}
\setcctype{by}
\acmConference[WWW '26] {Proceedings of the ACM Web Conference 2026}{April 13--17, 2026}{Dubai, United Arab Emirates.}
\acmBooktitle{Proceedings of the ACM Web Conference 2026 (WWW '26), April 13--17, 2026, Dubai, United Arab Emirates}
\acmISBN{979-8-4007-2307-0/2026/04}
\acmDOI{10.1145/3774904.3792960}

\begin{document}

\title{Agent-Enhanced Heterogeneous Graph RAG for Academic Question Answering}


\author{Runsong Jia}
\affiliation{%
  \institution{University of Technology Sydney}
  \city{Sydney}
  \country{Australia}}
\email{runsong.jia@student.uts.edu.au}

\author{Mengjia Wu}
\affiliation{%
  \institution{University of Technology Sydney}
  \city{Sydney}
  \country{Australia}}
\email{mengjia.wu@uts.edu.au}

\author{Ying Ding}
\affiliation{%
  \institution{University of Texas at Austin}
  \city{Austin}
  \country{United States}}
\email{ying.ding@ischool.utexas.edu}

\author{Jie Lu}
\affiliation{%
  \institution{University of Technology Sydney}
  \city{Sydney}
  \country{Australia}}
\email{jie.lu@uts.edu.au}

\author{Yi Zhang}
\authornote{Corresponding author.}
\affiliation{%
  \institution{University of Technology Sydney}
  \city{Sydney}
  \country{Australia}}
\email{yi.zhang@uts.edu.au}

\renewcommand{\shortauthors}{Runsong Jia, Mengjia Wu, Ying Ding, Jie Lu, and Yi Zhang}

\begin{abstract}
Academic question answering requires reasoning over heterogeneous scholarly graphs, where queries range from simple attribute lookups to multi-hop inference across author--paper--venue structures. Existing retrieval-augmented generation (RAG) systems struggle in this setting due to three limitations: (1) fixed retrieval strategies that do not adapt to varying query complexity, (2) the absence of sufficiency evaluation leading to incomplete or misaligned evidence, and (3) a lack of structured verification against graph facts. To address these issues, we propose an agentic heterogeneous graph RAG method that transforms the three core stages of the RAG pipeline into explicit agentic decision steps. A query-aware retrieval agent analyzes query type and selects an appropriate graph traversal strategy; a sufficiency-aware reranking agent assesses evidence completeness and adaptively expands the retrieved subgraph; and a graph-grounded verification agent checks entity, relation, and attribute correctness before finalizing the answer. Experiments on heterogeneous graphs constructed from OpenAlex and DBLP suggest that our method consistently outperforms strong LLM, graph-augmented RAG, and agent-based baselines.

\end{abstract}





\maketitle

\section{Introduction}

Academic question answering (QA) requires combining natural-language understanding with structured knowledge encoded in heterogeneous scholarly graphs. RAG has become a standard paradigm for knowledge-intensive QA \cite{lewis2020rag}, and recent studies explore combining LLMs with knowledge graphs or heterogeneous graphs to enhance multi-hop reasoning and factual grounding \cite{he2024gretriever}.

Despite these advances, current graph-augmented RAG methods still exhibit three limitations. They rarely model query complexity, applying uniform traversal depth to both simple attribute queries and multi-hop reasoning queries \cite{pan2024roadmap}. They lack evidence sufficiency control, typically retrieving a one-shot subgraph without ensuring that it satisfies the entities, constraints, or aggregation signals implied by the query. Finally, they generally do not perform graph-grounded answer verification: once evidence is retrieved, generated answers are seldom checked against the graph for entity existence, relation validity, or attribute correctness. In contrast, text-only hallucination detection approaches such as SelfCheckGPT \cite{manakul2023selfcheckgpt} and self-consistency methods \cite{wang2023selfconsistency} primarily rely on self-consistency rather than structured verification grounded in graph facts. This motivates our central question:

\begin{center}
\textbf{\textit{How can we develop an agentic RAG method for Academic QA in heterogeneous scholarly graphs?}}
\end{center}

We address this question by introducing an agent-enhanced heterogeneous graph RAG method in which the three stages of the RAG pipeline are explicitly agenticized. A query-aware retrieval agent analyzes the type of the question and produces a retrieval plan tailored to attribute, direct relation, aggregation, or multi-hop queries. A sufficiency-aware evidence control agent evaluates whether the retrieved subgraph adequately supports the query and selectively expands the graph neighborhood when necessary, rather than relying on fixed-depth traversal. A graph-grounded verification agent checks entity existence, relation validity, attribute correctness, and answer completeness before returning the final answer, inspired by recent agentic frameworks that decompose complex tasks into specialized modules \cite{yao2023react, singh2025agenticrag}. 

We evaluate the method on heterogeneous graphs constructed from OpenAlex and DBLP, comparing it with pure LLM baselines, representative graph-augmented RAG models, and agent-based retrieval methods. Across all evaluation settings, the agentic decomposition leads to consistent improvements in answer accuracy, demonstrating the effectiveness of integrating agentic query-aware planning, adaptive retrieval, and graph-grounded verification.

Our main contributions are summarized as follows:
\begin{enumerate}
\item We propose an agentic method for the three core stages of the Academic QA RAG pipeline: retrieval planning, evidence control, and answer verification, where each stage is handled by a dedicated agent that explicitly leverages heterogeneous scholarly graph structure.
\item We develop a query and sufficiency-aware graph retrieval method that classifies questions into four academic query types and adaptively guides retrieval and reranking based on query structure.
\item We propose a graph-grounded verification method that validates answers against retrieved graph evidence, which helps improve the robustness of the overall method.
\end{enumerate}

\section{Related Works}

\paragraph{\textbf{Academic QA and Heterogeneous Scholarly Graphs}}
Scholarly data—comprising authors, papers, venues, and citation relations—naturally forms heterogeneous graphs. Prior research in academic recommendation, citation analysis, and structured QA has leveraged graph neural networks (GNNs) and knowledge-graph reasoning to encode such structures. Representative methods, including Heterogeneous Graph Transformer (HGT) \cite{hu2020hgt}, learn type-aware representations by aggregating information across nodes and edges. While this line of work has advanced structured reasoning over academic graphs, most methods focus on embedding learning or executing logical operators, and relatively fewer studies explore integrating LLMs as generators or incorporating explicit mechanisms that adapt retrieval to the query’s structure or complexity.


\paragraph{\textbf{Retrieval-Augmented Generation and Graph-RAG}}
RAG has become a standard paradigm for knowledge-intensive QA, where retrieved evidence guides an LLM during generation \cite{lewis2020rag}. Extending this idea, graph- and KG-augmented RAG variants incorporate graph traversal into retrieval pipelines, enabling the use of structured relations as evidence \cite{jia2025hetgcot}. Recent systems such as G-Retriever \cite{he2024gretriever} retrieves subgraphs to support textual graph understanding and integrate graph foundation models into RAG to improve retrieval over structured relations. These graph RAG systems highlight the benefits of combining structured knowledge with LLMs for knowledge-intensive QA.


\paragraph{\textbf{Agentic LLMs and Answer Verification}}
Agentic LLM paradigms view language models as controllers capable of invoking tools, planning actions, or coordinating subtasks. ReAct \cite{yao2023react} demonstrates interleaved reasoning and acting, while MAIN-RAG \cite{chang2025mainrag} proposes a multi-agent filtering mechanism for retrieval-augmented generation. Parallel to this, answer verification and reliability assessment have received growing attention, with methods such as self-consistency \cite{wang2023selfconsistency} and FActScore \cite{min2023factscore} offering fine-grained evaluation over generated text. Recent efforts have also explored agent-based or complexity-aware RAG methods \cite{jeong2024adaptiverag} \cite{lee2024agentg}, which demonstrate the potential of using agents to control retrieval behavior. These approaches primarily operate in unstructured or text-centric settings, whereas our work focuses on verification against structured evidence from heterogeneous scholarly graphs.

\section{Methodology}

\begin{figure}[ht]
    \centering
    \includegraphics[width=\columnwidth]{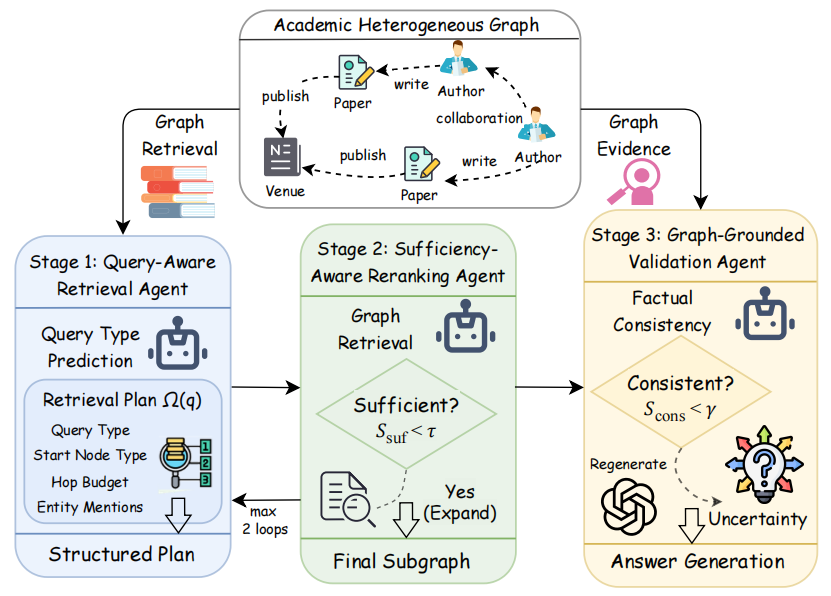}
    \caption{Overall architecture of our method.}
    \label{fig:architecture}
\end{figure}

This section introduces the academic heterogeneous graph setting and describes
each agent and its decision rules.
As shown in Figure~\ref{fig:architecture}, we propose an agentic heterogeneous graph RAG method that decomposes
Academic QA into three decision stages: (1) query-aware retrieval planning,
(2) graph retrieval with sufficiency-aware evidence control, and
(3) graph-grounded answer verification.

\subsection{Academic Heterogeneous Graph QA Setting}
We consider a heterogeneous scholarly graph
$\mathcal{G} = (\mathcal{V}, \mathcal{E})$
with three node types: Authors, Papers, and Venues.
Author nodes store attributes such as name and organization, Paper nodes
include title, abstract, keywords, citation counts, and year, and Venue
nodes include venue name and rank.
The graph includes two relation types: Paper--Author, representing authorship, 
and Paper--Venue, representing publication venues.

Given a natural-language query $q$, the goal is to generate an answer $a$
grounded in a task-specific subgraph $\mathcal{G}_q \subseteq \mathcal{G}$. Motivated by prior analyses of complex Knowledge Base QA and complex logical
queries over knowledge graphs ~\cite{lan2021complexkbqa}, we classify queries into four structural
categories:
Attribute (0-hop), Direct Relation (1-hop),
Aggregation (1--2 hops with numerical reasoning),
and Multi-hop (2+ hop relational chains).
Each type induces different requirements for start-node selection and
traversal depth.

\subsection{Query-Aware Retrieval Agent}

The Retrieval Agent converts a natural-language query into a
structured retrieval plan.
Specifically, given query $q$, the agent predicts:
(1) query type $t \in \{\mathrm{Attr}, \mathrm{Rel}, \mathrm{Agg}, \mathrm{Multi}\}$,
(2) start-node type $s \in \{\mathrm{Author}, \mathrm{Paper}, \mathrm{Venue}\}$,
(3) hop budget $h$ determined by $t$,
(4) entity mentions $E(q)$, and
(5) traversal plan $\pi_t$ (node lookup, single-hop, multi-hop, aggregation).

We formalize the plan as:
$\Omega(q) = (t, s, h, E(q), \pi_t)$. The agent uses an LLM classifier to infer $t$ with distribution:
\begin{equation}
    p(t \mid q) = \mathrm{softmax}\!\left(f_{\theta}(q)\right)
\end{equation}
The predicted query type is selected as:
\begin{equation}
    t^\star = \arg\max_{t} p(t \mid q)
\end{equation}
after which $(s, h, \pi_t)$ are determined through a fixed mapping.

In practice, $f_{\theta}$ is implemented as a prompted GPT-4-turbo
classifier: we provide a system prompt that describes the academic graph
schema and the four query types, together with a few in-context examples,
and let the model directly output a description of
$\Omega(q)$ without additional supervised fine-tuning.
This step makes retrieval planning explicit and agentic.

\subsection{Graph Retrieval and Sufficiency-Aware Reranking Agent}

Given $\Omega(q)$, the system performs (1) entity lookup, (2) hop constrained traversal, and (3) sufficiency-aware evidence control.
The start node is determined by embedding similarity:

\begin{equation}
v^\star = \arg\max_{v \in \mathcal{V}_s} \mathrm{sim}(\phi(q), \phi(v))
\end{equation}

where $\phi(\cdot)$ is a text encoder.
We instantiate $\phi$ using a Sentence-Transformers model (all-MiniLM-L6-v2),
which produces 384 dimensional embeddings suitable for efficient semantic similarity matching between queries and node metadata.
Traversal then yields the initial subgraph as $\mathcal{G}_q^{(0)} = \mathrm{Traverse}(v^\star, h, \pi_t)$.

The Reranking Agent evaluates whether $\mathcal{G}_q^{(i)}$ is adequate to
answer $q$ through a sufficiency score:
\begin{equation}
    S_{\text{suf}}(q, \mathcal{G}_q^{(i)}) =
    \alpha\, \mathrm{Cov}(q, \mathcal{G}_q^{(i)}) +
    (1-\alpha)\, \mathrm{Rel}(q, \mathcal{G}_q^{(i)})
\end{equation}
where $\mathrm{Cov}$ measures entity/attribute coverage and
$\mathrm{Rel}$ measures semantic relevance.
Concretely, $\mathrm{Cov}(q, \mathcal{G}_q^{(i)})$ is computed as the
fraction of entity mentions in $E(q)$ that can be matched to nodes in
$\mathcal{G}_q^{(i)}$ via string and embedding-based matching, while
$\mathrm{Rel}(q, \mathcal{G}_q^{(i)})$ averages the embedding similarity
between the query and textual descriptions of nodes in the subgraph. If the score is below threshold ($S_{\text{suf}} < \tau$),
the agent triggers controlled expansion:
\begin{equation}
    \mathcal{G}_q^{(i+1)} =
    \mathrm{Expand}(\mathcal{G}_q^{(i)}, \pi_t)
\end{equation}
with at most two expansions.
In our experiments, we cap expansion at two steps to control latency and keep the retrieved subgraph compact, introducing explicit evidence control into the RAG pipeline.

\subsection{Answer Generation and Graph-Grounded Validation Agent}

The retrieved subgraph $\mathcal{G}_q$ is linearized and supplied to an LLM
to generate answer $a$.
The Validation Agent then checks whether the answer aligns with graph facts
with factual consistency score defined as follows:
\begin{equation}
\begin{aligned}
S_{\text{cons}}(a, \mathcal{G}_q)
= {} & 
\beta_1\, \mathrm{EntAlign}(a, \mathcal{G}_q) 
+ \beta_2\, \mathrm{RelAlign}(a, \mathcal{G}_q) \\
& + \beta_3\, \mathrm{AttrMatch}(a, \mathcal{G}_q)
\end{aligned}
\end{equation}
where the three components evaluate entity existence, relation validity,
and attribute correctness.

Answer entities, relations, and attribute claims are extracted from $a$
using lightweight LLM-based tagging prompts.
$\mathrm{EntAlign}(a, \mathcal{G}_q)$ measures the fraction of entity
mentions in the answer that can be linked to nodes in $\mathcal{G}_q$
via string matching and embedding similarity; $\mathrm{RelAlign}(a,
\mathcal{G}_q)$ checks whether relations described in the answer
(e.g., head–relation–tail triples) correspond to existing edges in the
retrieved subgraph; and $\mathrm{AttrMatch}(a, \mathcal{G}_q)$ verifies
that numerical and categorical attributes in the answer (such as years,
venues, or counts) agree with the corresponding node attributes. 

If $S_{\text{cons}} < \gamma$, the agent attempts regeneration (up to two times) or outputs an uncertainty signal. We limit regeneration attempts to two to bound computation cost. This graph-grounded check helps detect unsupported or hallucinated content.

\section{Experimental Setup}

\paragraph{\textbf{Datasets}}
We evaluate our method on heterogeneous academic graphs derived from OpenAlex and DBLP. Each graph contains three node types (Authors, Papers, Venues) and two relations (writes, published\_in). OpenAlex includes 76{,}569 nodes and 105{,}290 edges, while DBLP contains 62{,}443 nodes and 79{,}697 edges. Following our query taxonomy, we generate 400 queries for each dataset (100 per type: attribute, direct-relation, aggregation, multi-hop), with ground-truth answers automatically extracted from graph facts.

\paragraph{\textbf{Backbone LLM and Implementation Details}}
All agentic controllers and the final answer generator use GPT-4-turbo. Entity lookup is performed using \texttt{sentence-transformers} for semantic similarity matching. For each query, the system allows at most two retrieval expansions and at most two verification-based regenerations. We report three standard metrics: (1) Accuracy, which measures exact match with the gold answer; (2) F1, computed via entity-level precision and recall; and (3) Hit@1, which evaluates whether the top-ranked predicted answer is correct.

\paragraph{\textbf{Baselines}}
We compare our method against three categories of baselines: (1) pure LLM models, including Qwen-2.5-7B and GPT-o3; (2) graph-augmented RAG methods, including Vanilla RAG, GraphPrompter ~\cite{liu2024graphprompter}, GraphRAG ~\cite{edge2024graphrag}, KGRAG ~\cite{sanmartin2024kgrag}, and GraphCoT ~\cite{jin2024graphcot}; and (3) agent-based or complexity-aware RAG methods, including AdaptiveRAG ~\cite{jeong2024adaptiverag} (agentic, complexity-aware retrieval) and Agent-G ~\cite{lee2024agentg} (agentic retrieval control). These baselines cover the major types of RAG, graph-RAG, and agent-based QA methods.

\section{Results}

\paragraph{\textbf{Overall Performance}}
Table~\ref{tab:results} shows that our method achieves the strongest performance on both OpenAlex and DBLP across all metrics. Compared with graph-augmented approaches such as GraphCoT, GraphRAG, and KGRAG, as well as agent-based methods like AdaptiveRAG and Agent-G, our method delivers clear performance improvements, reaching 76.68\% accuracy on OpenAlex and 73.43\% accuracy on DBLP. The improvements are particularly notable in Hit@1, indicating that query-aware planning and sufficiency-aware retrieval lead to more reliable answer selection.

\begin{figure}[ht]
    \centering
    \includegraphics[width=\columnwidth]{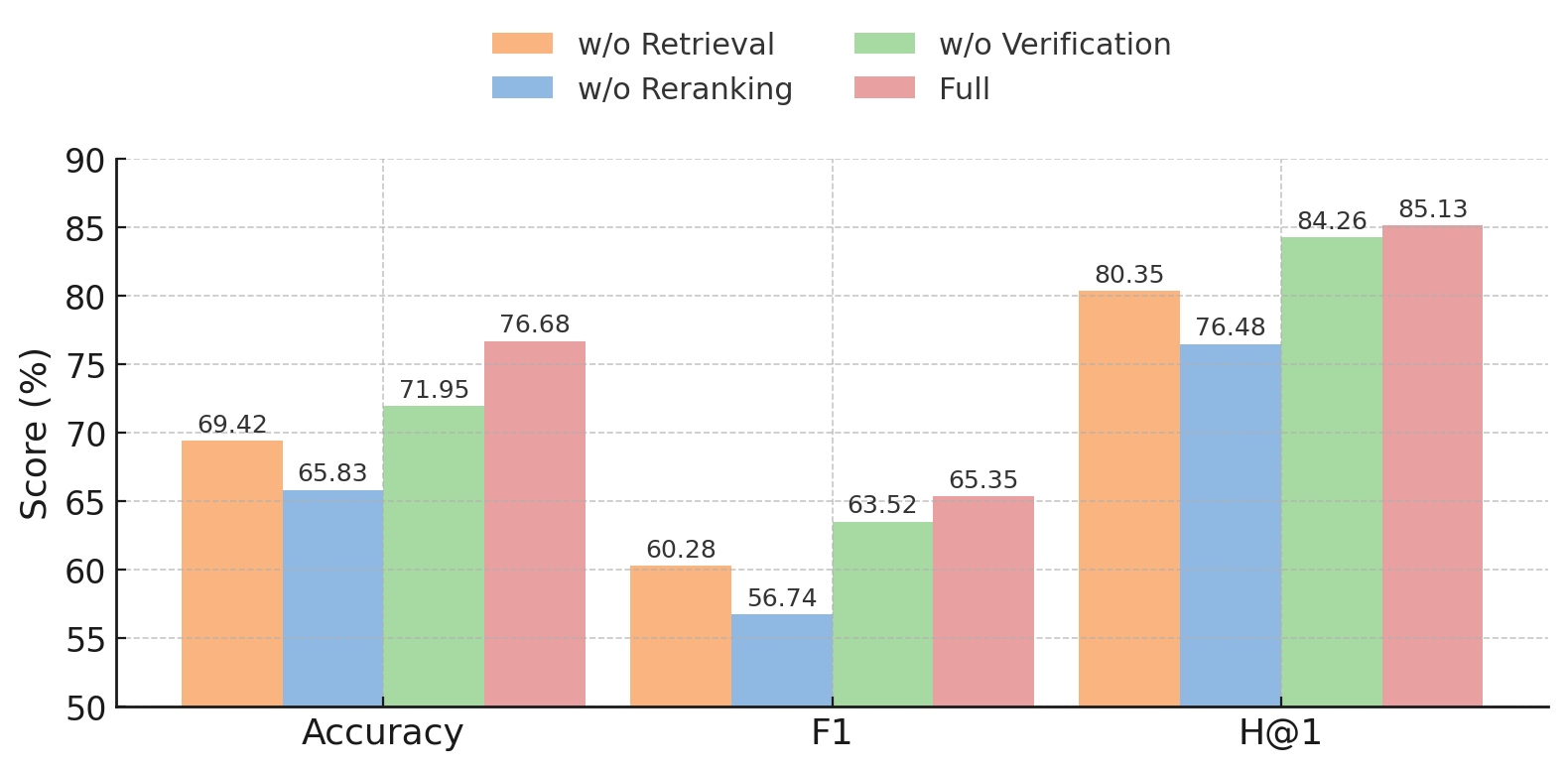}
    \caption{Ablation study on OpenAlex}
    \label{fig:openalex}
\end{figure}

\begin{figure}[ht]
    \centering
    \includegraphics[width=\columnwidth]{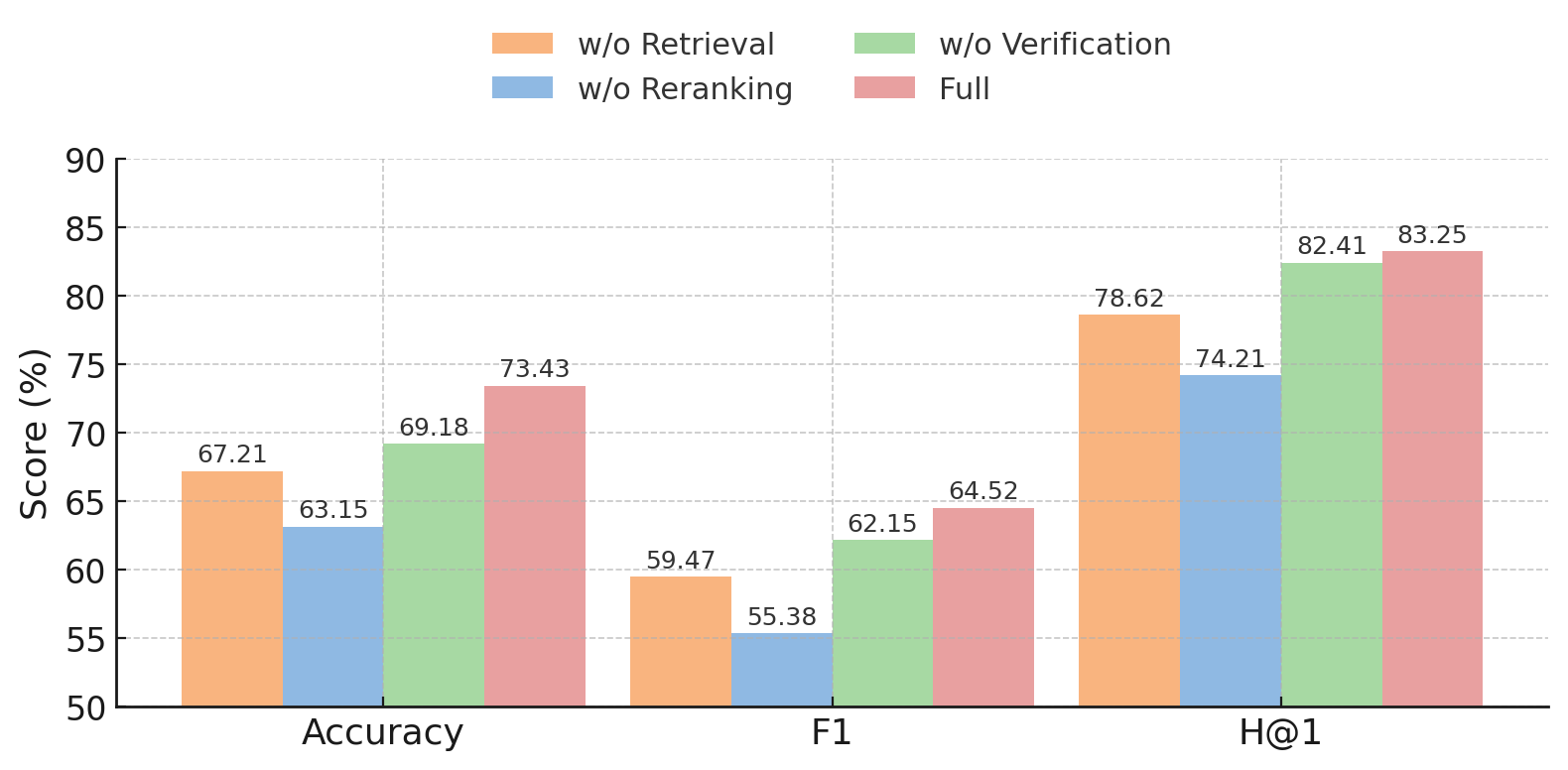}
    \caption{Ablation study on DBLP}
    \label{fig:dblp}
\end{figure}

\paragraph{\textbf{Ablation Study}}
Figure~\ref{fig:openalex} and figure~\ref{fig:dblp} report ablations for the three agents in our method. Removing the Retrieval Agent yields the largest performance drop, underscoring the importance of explicit query-type analysis and hop-budget planning. The Reranking Agent also contributes substantially by enabling adaptive evidence expansion when the initial subgraph is insufficient. Finally, the Verification Agent improves F1 and Hit@1 by checking entity, relation, and attribute correctness against the retrieved subgraph. Overall, each agent contributes meaningfully to the method, and the full agentic pipeline provides the most robust performance.

\begin{table}
  \caption{Performance comparison of different models on OpenAlex and DBLP.}
  \label{tab:results}
  \centering
  \resizebox{\linewidth}{!}{%
  \begin{tabular}{lcccccc}
    \toprule
    Model & \multicolumn{3}{c}{OpenAlex} & \multicolumn{3}{c}{DBLP} \\
    \cmidrule(lr){2-4} \cmidrule(lr){5-7}
          & Accuracy & F1 & H@1 & Accuracy & F1 & H@1 \\
    \midrule
    Qwen 2.5 7B      & 48.63 & 43.46 & 50.26 & 45.93 & 39.82 & 51.52 \\
    GPT o3           & 53.58 & 46.45 & 51.65 & 51.65 & 45.59 & 50.67 \\
    Vanilla RAG      & 54.36 & 48.92 & 63.27 & 52.14 & 46.82 & 65.43 \\
    GraphPrompter    & 59.28 & 52.46 & 62.23 & 57.63 & 49.25 & 67.21 \\
    GraphRAG         & 61.87 & 57.39 & 74.86 & 60.27 & 53.32 & 72.59 \\
    KGRAG            & 62.49 & 55.28 & 68.51 & 62.35 & 54.71 & 73.48 \\
    GraphCoT         & 64.28 & 56.43 & 75.62 & 68.52 & 59.43 & 78.26 \\
    AdaptiveRAG      & 72.58 & 62.95 & 82.63 & 70.23 & 61.29 & 80.48 \\
    Agent-G          & 73.14 & 61.74 & 81.15 & 69.74 & 60.82 & 79.86 \\
    Ours             & \textbf{76.68} & \textbf{65.35} & \textbf{85.13}
                     & \textbf{73.43} & \textbf{64.52} & \textbf{83.25} \\
    \bottomrule
  \end{tabular}
  }%
\end{table}

\vspace{-10pt}   


\section{Conclusion}

In conclusion, we introduced an agentic heterogeneous graph RAG method for Academic QA that incorporates query-aware retrieval planning, sufficiency-aware evidence control, and graph-grounded answer verification. By explicitly agenticizing the three major steps of the RAG pipeline, our method more effectively aligns retrieval and verification with the structure of heterogeneous scholarly graphs. Experiments on OpenAlex and DBLP demonstrate consistent improvements over pure LLM baselines, graph-augmented RAG systems, and agent-based retrieval methods.
\section*{Acknowledgments}
This work was supported by the Australian Commonwealth Scientific and Industrial Research Organization (CSIRO) in conjunction with the National Science Foundation (NSF) of the United States, under CSIRO-NSF \#2303037.

\bibliographystyle{ACM-Reference-Format}
\balance
\bibliography{sample-base}


\begin{thebibliography}{18}


\ifx \showCODEN    \undefined \def \showCODEN     #1{\unskip}     \fi
\ifx \showISBNx    \undefined \def \showISBNx     #1{\unskip}     \fi
\ifx \showISBNxiii \undefined \def \showISBNxiii  #1{\unskip}     \fi
\ifx \showISSN     \undefined \def \showISSN      #1{\unskip}     \fi
\ifx \showLCCN     \undefined \def \showLCCN      #1{\unskip}     \fi
\ifx \shownote     \undefined \def \shownote      #1{#1}          \fi
\ifx \showarticletitle \undefined \def \showarticletitle #1{#1}   \fi
\ifx \showURL      \undefined \def \showURL       {\relax}        \fi
\providecommand\bibfield[2]{#2}
\providecommand\bibinfo[2]{#2}
\providecommand\natexlab[1]{#1}
\providecommand\showeprint[2][]{arXiv:#2}

\bibitem[Chang et~al\mbox{.}(2025)]%
        {chang2025mainrag}
\bibfield{author}{\bibinfo{person}{Chia-Yuan Chang}, \bibinfo{person}{Zhimeng Jiang}, \bibinfo{person}{Vineeth Rakesh}, \bibinfo{person}{Menghai Pan}, \bibinfo{person}{Chin-Chia~Michael Yeh}, \bibinfo{person}{Guanchu Wang}, \bibinfo{person}{Mingzhi Hu}, \bibinfo{person}{Zhichao Xu}, \bibinfo{person}{Yan Zheng}, \bibinfo{person}{Mahashweta Das}, {and} \bibinfo{person}{Na Zou}.} \bibinfo{year}{2025}\natexlab{}.
\newblock \showarticletitle{Main-rag: Multi-agent filtering retrieval-augmented generation}. In \bibinfo{booktitle}{\emph{Proceedings of the 63rd Annual Meeting of the Association for Computational Linguistics (Volume 1: Long Papers)}}. \bibinfo{pages}{2607--2622}.
\newblock


\bibitem[Edge et~al\mbox{.}(2024)]%
        {edge2024graphrag}
\bibfield{author}{\bibinfo{person}{Darren Edge}, \bibinfo{person}{Ha Trinh}, \bibinfo{person}{Newman Cheng}, \bibinfo{person}{Joshua Bradley}, \bibinfo{person}{Alex Chao}, \bibinfo{person}{Apurva Mody}, \bibinfo{person}{Steven Truitt}, \bibinfo{person}{Dasha Metropolitansky}, \bibinfo{person}{Robert~Osazuwa Ness}, {and} \bibinfo{person}{Jonathan Larson}.} \bibinfo{year}{2024}\natexlab{}.
\newblock \showarticletitle{From local to global: A graph rag approach to query-focused summarization}.
\newblock \bibinfo{journal}{\emph{arXiv preprint arXiv:2404.16130}} (\bibinfo{year}{2024}).
\newblock


\bibitem[He et~al\mbox{.}(2024)]%
        {he2024gretriever}
\bibfield{author}{\bibinfo{person}{Xiaoxin He}, \bibinfo{person}{Yijun Tian}, \bibinfo{person}{Yifei Sun}, \bibinfo{person}{Nitesh Chawla}, \bibinfo{person}{Thomas Laurent}, \bibinfo{person}{Yann LeCun}, \bibinfo{person}{Xavier Bresson}, {and} \bibinfo{person}{Bryan Hooi}.} \bibinfo{year}{2024}\natexlab{}.
\newblock \showarticletitle{G-retriever: Retrieval-augmented generation for textual graph understanding and question answering}.
\newblock \bibinfo{journal}{\emph{Advances in Neural Information Processing Systems}}  \bibinfo{volume}{37} (\bibinfo{year}{2024}), \bibinfo{pages}{132876--132907}.
\newblock


\bibitem[Hu et~al\mbox{.}(2020)]%
        {hu2020hgt}
\bibfield{author}{\bibinfo{person}{Ziniu Hu}, \bibinfo{person}{Yuxiao Dong}, \bibinfo{person}{Kuansan Wang}, {and} \bibinfo{person}{Yizhou Sun}.} \bibinfo{year}{2020}\natexlab{}.
\newblock \showarticletitle{Heterogeneous graph transformer}. In \bibinfo{booktitle}{\emph{Proceedings of the web conference 2020}}. \bibinfo{pages}{2704--2710}.
\newblock


\bibitem[Jeong et~al\mbox{.}(2024)]%
        {jeong2024adaptiverag}
\bibfield{author}{\bibinfo{person}{Soyeong Jeong}, \bibinfo{person}{Jinheon Baek}, \bibinfo{person}{Sukmin Cho}, \bibinfo{person}{Sung~Ju Hwang}, {and} \bibinfo{person}{Jong~C Park}.} \bibinfo{year}{2024}\natexlab{}.
\newblock \showarticletitle{Adaptive-RAG: Learning to Adapt Retrieval-Augmented Large Language Models through Question Complexity}. In \bibinfo{booktitle}{\emph{Proceedings of the 2024 Conference of the North American Chapter of the Association for Computational Linguistics: Human Language Technologies (Volume 1: Long Papers)}}. \bibinfo{pages}{7029--7043}.
\newblock


\bibitem[Jia et~al\mbox{.}(2025)]%
        {jia2025hetgcot}
\bibfield{author}{\bibinfo{person}{Runsong Jia}, \bibinfo{person}{Mengjia Wu}, \bibinfo{person}{Ying Ding}, \bibinfo{person}{Jie Lu}, {and} \bibinfo{person}{Yi Zhang}.} \bibinfo{year}{2025}\natexlab{}.
\newblock \showarticletitle{HetGCoT: Heterogeneous Graph-Enhanced Chain-of-Thought LLM Reasoning for Academic Question Answering}. In \bibinfo{booktitle}{\emph{Findings of the Association for Computational Linguistics: EMNLP 2025}}. \bibinfo{publisher}{Association for Computational Linguistics}, \bibinfo{pages}{15950--15963}.
\newblock
\href{https://doi.org/10.18653/v1/2025.findings-emnlp.864}{doi:\nolinkurl{10.18653/v1/2025.findings-emnlp.864}}


\bibitem[Jin et~al\mbox{.}(2024)]%
        {jin2024graphcot}
\bibfield{author}{\bibinfo{person}{Bowen Jin}, \bibinfo{person}{Chulin Xie}, \bibinfo{person}{Jiawei Zhang}, \bibinfo{person}{Kashob~Kumar Roy}, \bibinfo{person}{Yu Zhang}, \bibinfo{person}{Zheng Li}, \bibinfo{person}{Ruirui Li}, \bibinfo{person}{Xianfeng Tang}, \bibinfo{person}{Suhang Wang}, \bibinfo{person}{Yu Meng}, {et~al\mbox{.}}} \bibinfo{year}{2024}\natexlab{}.
\newblock \showarticletitle{Graph Chain-of-Thought: Augmenting Large Language Models by Reasoning on Graphs}. In \bibinfo{booktitle}{\emph{Findings of the Association for Computational Linguistics ACL 2024}}. \bibinfo{pages}{163--184}.
\newblock


\bibitem[Lan et~al\mbox{.}(2022)]%
        {lan2021complexkbqa}
\bibfield{author}{\bibinfo{person}{Yunshi Lan}, \bibinfo{person}{Gaole He}, \bibinfo{person}{Jinhao Jiang}, \bibinfo{person}{Jing Jiang}, \bibinfo{person}{Wayne~Xin Zhao}, {and} \bibinfo{person}{Ji-Rong Wen}.} \bibinfo{year}{2022}\natexlab{}.
\newblock \showarticletitle{Complex knowledge base question answering: A survey}.
\newblock \bibinfo{journal}{\emph{IEEE Transactions on Knowledge and Data Engineering}} \bibinfo{volume}{35}, \bibinfo{number}{11} (\bibinfo{year}{2022}), \bibinfo{pages}{11196--11215}.
\newblock


\bibitem[Lee et~al\mbox{.}(2024)]%
        {lee2024agentg}
\bibfield{author}{\bibinfo{person}{Meng-Chieh Lee}, \bibinfo{person}{Qi Zhu}, \bibinfo{person}{Costas Mavromatis}, \bibinfo{person}{Zhen Han}, \bibinfo{person}{Soji Adeshina}, \bibinfo{person}{Vassilis~N. Ioannidis}, \bibinfo{person}{Huzefa Rangwala}, {and} \bibinfo{person}{Christos Faloutsos}.} \bibinfo{year}{2024}\natexlab{}.
\newblock \showarticletitle{Agent-G: An Agentic Framework for Graph Retrieval Augmented Generation}.
\newblock  (\bibinfo{year}{2024}).
\newblock
\urldef\tempurl%
\url{https://openreview.net/forum?id=g2C947jjjQ}
\showURL{%
\tempurl}
\newblock
\shownote{Openreview Preprint}.


\bibitem[Lewis et~al\mbox{.}(2020)]%
        {lewis2020rag}
\bibfield{author}{\bibinfo{person}{Patrick Lewis}, \bibinfo{person}{Ethan Perez}, \bibinfo{person}{Aleksandra Piktus}, \bibinfo{person}{Fabio Petroni}, \bibinfo{person}{Vladimir Karpukhin}, \bibinfo{person}{Naman Goyal}, \bibinfo{person}{Heinrich K{\"u}ttler}, \bibinfo{person}{Mike Lewis}, \bibinfo{person}{Wen-tau Yih}, \bibinfo{person}{Tim Rockt{\"a}schel}, {et~al\mbox{.}}} \bibinfo{year}{2020}\natexlab{}.
\newblock \showarticletitle{Retrieval-augmented generation for knowledge-intensive nlp tasks}.
\newblock \bibinfo{journal}{\emph{Advances in Neural Information Processing Systems}}  \bibinfo{volume}{33} (\bibinfo{year}{2020}), \bibinfo{pages}{9459--9474}.
\newblock


\bibitem[Liu et~al\mbox{.}(2024)]%
        {liu2024graphprompter}
\bibfield{author}{\bibinfo{person}{Zheyuan Liu}, \bibinfo{person}{Xiaoxin He}, \bibinfo{person}{Yijun Tian}, {and} \bibinfo{person}{Nitesh~V Chawla}.} \bibinfo{year}{2024}\natexlab{}.
\newblock \showarticletitle{Can we soft prompt llms for graph learning tasks?}. In \bibinfo{booktitle}{\emph{Companion Proceedings of the ACM Web Conference 2024}}. \bibinfo{pages}{481--484}.
\newblock


\bibitem[Manakul et~al\mbox{.}(2023)]%
        {manakul2023selfcheckgpt}
\bibfield{author}{\bibinfo{person}{Potsawee Manakul}, \bibinfo{person}{Adian Liusie}, {and} \bibinfo{person}{Mark Gales}.} \bibinfo{year}{2023}\natexlab{}.
\newblock \showarticletitle{Selfcheckgpt: Zero-resource black-box hallucination detection for generative large language models}. In \bibinfo{booktitle}{\emph{Proceedings of the 2023 Conference on Empirical Methods in Natural Language Processing}}. \bibinfo{pages}{9004--9017}.
\newblock


\bibitem[Min et~al\mbox{.}(2023)]%
        {min2023factscore}
\bibfield{author}{\bibinfo{person}{Sewon Min}, \bibinfo{person}{Kalpesh Krishna}, \bibinfo{person}{Xinxi Lyu}, \bibinfo{person}{Mike Lewis}, \bibinfo{person}{Wen-tau Yih}, \bibinfo{person}{Pang Koh}, \bibinfo{person}{Mohit Iyyer}, \bibinfo{person}{Luke Zettlemoyer}, {and} \bibinfo{person}{Hannaneh Hajishirzi}.} \bibinfo{year}{2023}\natexlab{}.
\newblock \showarticletitle{Factscore: Fine-grained atomic evaluation of factual precision in long form text generation}. In \bibinfo{booktitle}{\emph{Proceedings of the 2023 Conference on Empirical Methods in Natural Language Processing}}. \bibinfo{pages}{12076--12100}.
\newblock


\bibitem[Pan et~al\mbox{.}(2024)]%
        {pan2024roadmap}
\bibfield{author}{\bibinfo{person}{Shirui Pan}, \bibinfo{person}{Linhao Luo}, \bibinfo{person}{Yufei Wang}, \bibinfo{person}{Chen Chen}, \bibinfo{person}{Jiapu Wang}, {and} \bibinfo{person}{Xindong Wu}.} \bibinfo{year}{2024}\natexlab{}.
\newblock \showarticletitle{Unifying large language models and knowledge graphs: A roadmap}.
\newblock \bibinfo{journal}{\emph{IEEE Transactions on Knowledge and Data Engineering}} \bibinfo{volume}{36}, \bibinfo{number}{7} (\bibinfo{year}{2024}), \bibinfo{pages}{3580--3599}.
\newblock


\bibitem[Sanmartin(2024)]%
        {sanmartin2024kgrag}
\bibfield{author}{\bibinfo{person}{Diego Sanmartin}.} \bibinfo{year}{2024}\natexlab{}.
\newblock \showarticletitle{Kg-rag: Bridging the gap between knowledge and creativity}.
\newblock \bibinfo{journal}{\emph{arXiv preprint arXiv:2405.12035}} (\bibinfo{year}{2024}).
\newblock


\bibitem[Singh et~al\mbox{.}(2025)]%
        {singh2025agenticrag}
\bibfield{author}{\bibinfo{person}{Aditi Singh}, \bibinfo{person}{Abul Ehtesham}, \bibinfo{person}{Saket Kumar}, {and} \bibinfo{person}{Tala~Talaei Khoei}.} \bibinfo{year}{2025}\natexlab{}.
\newblock \showarticletitle{Agentic retrieval-augmented generation: A survey on agentic rag}.
\newblock \bibinfo{journal}{\emph{arXiv preprint arXiv:2501.09136}} (\bibinfo{year}{2025}).
\newblock


\bibitem[Wang et~al\mbox{.}(2023)]%
        {wang2023selfconsistency}
\bibfield{author}{\bibinfo{person}{Xuezhi Wang}, \bibinfo{person}{Jason Wei}, \bibinfo{person}{Dale Schuurmans}, \bibinfo{person}{Quoc Le}, \bibinfo{person}{Ed Chi}, \bibinfo{person}{Sharan Narang}, \bibinfo{person}{Aakanksha Chowdhery}, {and} \bibinfo{person}{Denny Zhou}.} \bibinfo{year}{2023}\natexlab{}.
\newblock \showarticletitle{Self-consistency improves chain of thought reasoning in language models}.
\newblock \bibinfo{journal}{\emph{International Conference on Learning Representations}} (\bibinfo{year}{2023}).
\newblock


\bibitem[Yao et~al\mbox{.}(2023)]%
        {yao2023react}
\bibfield{author}{\bibinfo{person}{S. Yao}, \bibinfo{person}{J. Zhao}, \bibinfo{person}{D. Yu}, \bibinfo{person}{N. Du}, \bibinfo{person}{I. Shafran}, \bibinfo{person}{K. Narasimhan}, {and} \bibinfo{person}{Y. Cao}.} \bibinfo{year}{2023}\natexlab{}.
\newblock \showarticletitle{ReAct: Synergizing reasoning and acting in language models}. In \bibinfo{booktitle}{\emph{Proceedings of the Eleventh International Conference on Learning Representations (ICLR)}}.
\newblock


\end{thebibliography}

\end{document}